# Image-Domain Material Decomposition for Dual-energy CT using Unsupervised Learning with Data-fidelity Loss

Junbo Peng[1], Chih-Wei Chang[1], Huiqiao Xie[2], Richard L.J. Qiu[1], Justin Roper[1], Tonghe Wang[2], Beth Bradshaw[1], Xiangyang Tang[3] and Xiaofeng Yang[1]*

[1]Department of Radiation Oncology and Winship Cancer Institute, Emory University, Atlanta, GA 30322, USA

[2]Department of Medical Physics, Memorial Sloan Kettering Cancer Center, New York, NY, 10065

[3]Department of Radiology and Imaging Sciences and Winship Cancer Institute, Emory University, Atlanta, GA 30322, USA

*Corresponding to: xiaofeng.yang@emory.edu

## Abstract

**Background:** Dual-energy CT (DECT) and material decomposition play vital roles in quantitative medical imaging. However, the decomposition process may suffer from significant noise amplification, leading to severely degraded image signal-to-noise ratios (SNRs). While existing iterative algorithms perform noise suppression using different image priors, these heuristic image priors cannot accurately represent the features of the target image manifold. Although deep learning-based decomposition methods have been reported, these methods are in the supervised-learning framework requiring paired data for training, which is not readily available in clinical settings.

**Purpose:** This work aims to develop an unsupervised-learning framework with data-measurement consistency for image-domain material decomposition in DECT.

**Methods:** The proposed framework combines iterative decomposition and deep learning-based image prior in a generative adversarial network (GAN) architecture. In the generator module, a data-fidelity loss is introduced to enforce the measurement consistency in material decomposition. In the discriminator module, the discriminator is trained to differentiate the low-noise material-specific images from the high-noise images. In this scheme, paired images of DECT and ground-truth material-specific images are not required for the model training. Once trained, the generator can perform image-domain material decomposition with noise suppression in a single step.

**Results:** In the simulation studies of head and lung digital phantoms, the proposed method reduced the noise in images by more than 97% and 91%, respectively. It also generated decomposed images with structural similarity index measures (SSIMs) greater than 0.95 against the ground truth. In the clinical head and lung patient studies, the proposed method suppressed the noise by more than 95% and 93%, respectively.

**Conclusions:** Since the invention of DECT, the noise amplification during material decomposition has been one of the biggest challenges, impeding its quantitative use in clinical practice. The proposed method performs accurate material decomposition with efficient noise suppression, eliminating the need for higher-dose DECT scans. Furthermore, the proposed method is within an unsupervised-learning framework, which does not require paired data for model training and resolves the issue of lack of ground-truth data in clinical scenarios.

# 1. Introduction

Dual-energy computed tomography (DECT) is an advanced medical imaging technology that plays a vital role in clinical practice for its superior capability of material differentiation by scanning a patient using two different energy spectra.[1] With acquired dual-energy projection data, material decomposition can decompose the DECT images to basis material images by exploring the attenuation properties of different materials over X-ray energy. Over the past decades, DECT and material decomposition have been widely used in many applications, including iodine quantification,[2] separation of calcification and contrast-agent in angiography,[3,4] pseudo-monochromatic image generation,[5,6] and virtual non-contrast imaging.[7,8]

The principle of DECT material decomposition is since there are two primary interactions between X-ray photons and matter, i.e., the photoelectric absorption and Compton scattering, in the diagnostic energy range. Thus, the X-ray attenuation in any material can be decomposed into two known energy-dependent functions corresponding to the two interactions or the linear attenuation coefficient of two basis materials. Theoretically, the material decomposition can be performed in either the projection or image domain. The projection-domain material decomposition is proposed in the original concept of DECT and has the advantage of beam hardening correction by incorporating the X-ray's polychromatism into the nonlinear forward model. The implementation of projection-domain decomposition can be divided into two-step and one-step methods. In the two-step strategy, the sinogram of two basis materials is nonlinearly decomposed from the projection data in the first step, from which the material-specific maps can be linearly reconstructed through analytical or iterative algorithms.[1] However, the first step necessitates that dual-energy measurements be geometrically aligned for each X-ray through the object, which cannot be fulfilled in many scanners, e.g., the dual-source and fast kVp-switching DECT. The one-step approach addresses the limitation of ray-alignment in data acquisition by simultaneously performing the decomposition and reconstruction in a single nonlinear optimization framework.[9,10] Nevertheless, due to the inherent nonconvexity of the optimization problem, the one-step approaches tend to be computationally intensive, and may exhibit slow convergence, be sensitive to initialization, and prone to get trapped in local minima, making them less robust for clinical applications.[11]

Different from the nonlinear decomposition model employed in the projection-domain methods, the image-domain decomposition operates directly on reconstructed DECT images via linear approximation. This method is more straightforward to implement but lacks the capability of beam hardening correction.[12-14] Compared to the projection-domain methods, the image-domain decomposition is much easier to implement, and has been implemented in commercial, especially the dual-source-dual-detector DECT scanners.[15] However, the direct material decomposition via matrix inversion suffers from significant noise magnification and severely degraded signal-to-noise ratios (SNRs).[12] This problem arises largely because of the overlap of X-ray spectra at different peak tube voltages and correlation of basis materials' attenuating property in the diagnostic energy range, which is usually manifested as a large condition number of the linear system.[12,16-18] The noise amplification problem can be alleviated by traditional image post-processing algorithms, e.g., filtering and smoothing-based methods at the cost of degraded spatial resolution.[19,20] To fully explore the statistical properties of material decomposition, the iterative image-domain decomposition methods progressively suppress noise during the decomposition to obtain low-noise material-specific images that are optimal in statistical sense. This strategy combines noise suppression and material decomposition into a single optimization problem by incorporating various image-domain prior information as the regularization terms, such as total variation (TV),[14,16,21] quadratic smoothness penalty,[12] edge-preserving regularization,[22] and similarity-based regularization.[23] While the iterative methods outperform the traditional image post-processing methods, they show limited capability in modeling the complex noise correlation in material decomposition, alter the image noise power spectrum (NPS) and thus image textures.[24,25] This is because the heuristic image priors cannot represent all the features of target image manifold but only depict some specific image features, e.g., the piece-wise constant property. Furthermore, the optimization framework in the iterative decomposition methods is time-consuming and computation-intensive, hindering its practice in clinical applications.[11]

With the emergence and success of deep learning techniques in medical imaging and image analysis, attempts have been made to perform DECT reconstruction and material decomposition. Compared to the explicit image prior employed in the iterative decomposition methods, the neural network can, at superior performance and higher computation efficiency, learn the material image distribution better in an implicit form and perform direct distribution transfer from the DECT images or sinograms to decomposed material-specific images. In the image domain, Zhang et al designed a butterfly-like network to perform the dual-material decomposition,[24] and Gong et al proposed an Incep-Net for three-material decomposition using multi-branch modules.[26] In the projection domain, Shi et al developed a modified U-net to acquire material-specific projections,[27] and Jiang et al used a ResNet to generate the projections of three basis materials from triple-energy data.[28] To fully utilize the information stored in different domains, Su *et al* proposed a DIRECT-Net to improve the dual-material decomposition performance using mutual-domain data,[11] and Zhu et al used a cross-domain network to perform three-material decomposition from dual-energy data.[29] Despite the demonstrated capabilities of these methods, all of them are in the supervised learning framework and require large-scale paired data of dual-energy or multi-energy CT and ground-truth material-specific images for model training, which is not always available in clinical practice.

In this work, we aim to develop an image-domain material decomposition method in an unsupervised-learning framework to tackle the challenges associated with the unavailability of ground-truth data in clinical applications. Inspired by the discrimination mechanism of the generative adversarial network (GAN),[30] we trained a discriminator network to capture the distribution of low-noise material-specific images and differentiate different noise levels of decomposed images. To guarantee the accuracy of decomposed images without the supervision of image labels, a data-fidelity loss is introduced to enforce the image-domain data consistency between the acquired and calculated DECT images. The feasibility of our method was evaluated and validated using digital phantom and clinical DECT data.

## 2. Methodology

### 2.1 Principles of image-domain material decomposition

In the image-domain decomposition of DECT, the linear attenuation coefficients ($\mu$) of each pixel pair in DECT images are approximated as a linear combination of the pixel values in basis material maps, i.e.,

$$\begin{bmatrix} \mu_H \\ \mu_L \end{bmatrix} = \begin{bmatrix} \mu_{1,H} & \mu_{2,H} \\ \mu_{1,L} & \mu_{2,L} \end{bmatrix} \begin{bmatrix} x_1 \\ x_2 \end{bmatrix} \quad (1)$$

where the subscript H/L denotes the high/low-energy spectrum and the index 1/2 represents different basis materials. The unit of length$^{-1}$ is used here, instead of the Hounsfield unit (HU), for the variable $\mu$ to preserve the system's linearity. The vector $[x_1, x_2]^\text{T}$ denotes the volume fraction of two basis materials, which are unitless.

To process the whole DECT images, Eq. (1) can be rewritten in the matrix-vector multiplication form as

$$\begin{bmatrix} \vec{\mu}_H \\ \vec{\mu}_L \end{bmatrix} = A \begin{bmatrix} \vec{x}_1 \\ \vec{x}_2 \end{bmatrix} \quad (2)$$

In the above equation, $\vec{\mu}_H$ and $\vec{\mu}_L$ are the measured high- and low-energy CT images represented in column vectors, respectively. Similarly, $\vec{x}_1$ and $\vec{x}_2$ are the vectorized material-specific images. $A$ is the material composition matrix, with a dimension of 2$N$-by-2$N$ and $N$ is the pixel number in one CT image. $A$ can be obtained from Eq. (1) as

$$A = \begin{bmatrix} \mu_{1,H}I & \mu_{2,H}I \\ \mu_{1,L}I & \mu_{2,L}I \end{bmatrix} \tag{3}$$

where $I$ is an identity matrix with a dimension of $N$-by-$N$.

Based on Eq. (3), one can directly calculate the decomposed material-specific images via matrix inversion as

$$\begin{bmatrix} \vec{x}_1 \\ \vec{x}_2 \end{bmatrix} = A^{-1} \begin{bmatrix} \vec{\mu}_H \\ \vec{\mu}_L \end{bmatrix} \tag{4}$$

where $A^{-1}$ is the decomposition matrix calculated by

$$A^{-1} = \frac{1}{\mu_{1,H}\mu_{2,L} - \mu_{2,H}\mu_{1,L}} \begin{bmatrix} \mu_{2,L}I & -\mu_{2,H}I \\ -\mu_{1,L}I & \mu_{1,H}I \end{bmatrix} \tag{5}$$

Noticeably, the image-domain method simplifies the material decomposition by treating it as a linear system, disregarding the polychromatic nature of X-ray beams. As a result, the implementation of material decomposition becomes more efficient, but at the cost of sacrificing the ability to correct beam-hardening effects.

**2.2 Iterative image-domain material decomposition with image regularization**

Direct decomposition using Eq. (4) is highly sensitive to noise and may produce images of basis materials with severely degraded SNRs compared to those of raw DECT images, caused by the poor condition of matrix $A$. To suppress the noise amplification during the image-domain material decomposition, the regularization-based iterative decomposition methods have been developed, which can be written as the following optimization problem

$$\min_{\vec{x}_1, \vec{x}_2} \frac{1}{2} \left\| A \begin{bmatrix} \vec{x}_1 \\ \vec{x}_2 \end{bmatrix} - \begin{bmatrix} \vec{\mu}_H \\ \vec{\mu}_L \end{bmatrix} \right\|_2^2 + \lambda_1 R(\vec{x}_1) + \lambda_2 R(\vec{x}_2) \tag{6}$$

where $R(\cdot)$ is the prior knowledge-based regularization function, e.g., TV or edge-preserving smoothness penalty, and $\lambda_{1,2}$ are user-defined parameters to balance the tradeoff between decomposition accuracy and noise suppression.

**2.3 Unsupervised-learning material decomposition with DECT data fidelity**

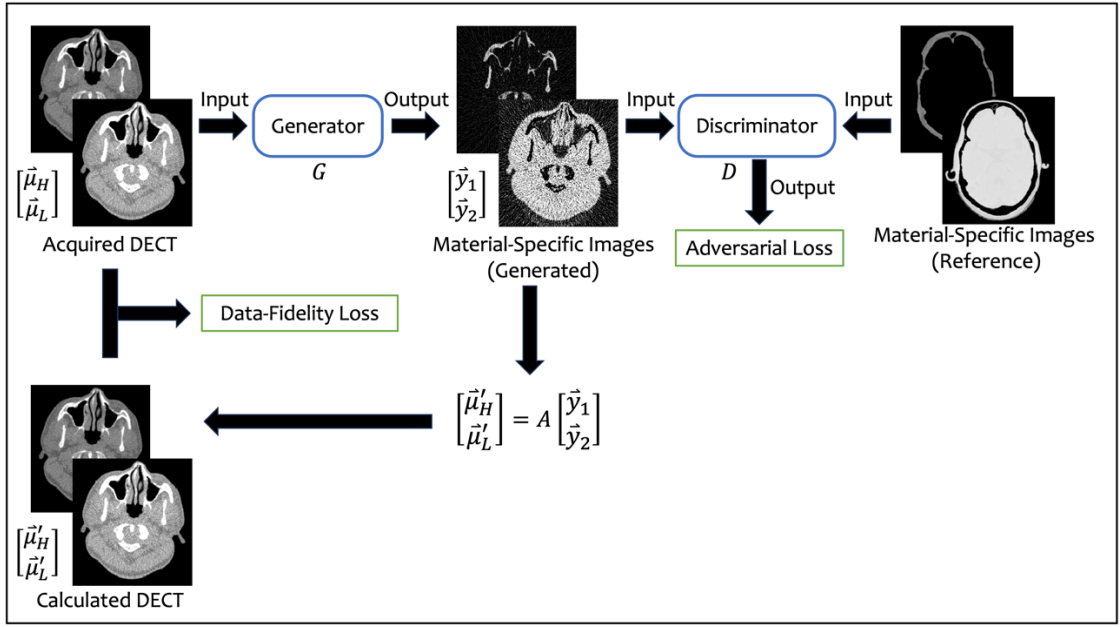

**Figure 1**. Workflow of the proposed unsupervised material decomposition method.

The proposed unsupervised framework for DECT material decomposition is depicted in Figure 1. In brief, the proposed method combines the principles of regularized iterative decomposition and the manifold learning capability of the neural network in a generative adversarial network (GAN) framework. There are two modules in the unsupervised learning framework: a generator network ($G$) for image-domain material decomposition and a discriminator network ($D$) for noise recognition. In the first module, concatenated DECT images are fed into the generator that outputs the material-specific images. Different from previously reported material decomposition networks minimizing the distance between the generated and ground-truth decomposed images, the proposed method minimizes the data discrepancy between the acquired DECT and calculated DECT from the generated material maps. Such a loss function is inspired by the DECT data fidelity term in Eq. (6), and is formulated as

$$Loss_{Data-Fidelity} = \left\| A \begin{bmatrix} \vec{y}_1 \\ \vec{y}_2 \end{bmatrix} - \begin{bmatrix} \vec{\mu}_H \\ \vec{\mu}_L \end{bmatrix} \right\|_2^2 \tag{7}$$

where $\vec{y}_{1,2}$ are the generated material-specific images from $G$, which can be formulated as

$$\begin{bmatrix} \vec{y}_1 \\ \vec{y}_2 \end{bmatrix} = G(\begin{bmatrix} \vec{\mu}_H \\ \vec{\mu}_L \end{bmatrix}) \tag{8}$$

In this way, the generator can learn an accurate mapping from the DECT images to decomposed images without the guidance of ground-truth material-specific images. This is the origin of the unsupervised-learning ability of the proposed material decomposition method.

In the second module, the discriminator network ($D$) is used to distinguish the high-noise decomposed images from the reference low-noise material-specific images. In other words, the discriminator aims to learn the manifold of low-noise images in an implicit form, providing the guidance for output of generator ($G$) in a way similar to the regularization term $R(\cdot)$ in Eq. 6. Compared to the artificially designed image priors in iterative methods, the implicitly learned feature by the discriminator is more general and accurate to represent the manifold of low-noise images. Besides, the reference images do not have to be associated with the input DECT. Thus, the proposed model can be trained in an unsupervised manner with no need for paired data. The total loss is the sum of data-fidelity loss and the conventional GAN loss,

$$Loss_{Total} = \lambda \cdot Loss_{Data-Fidelity} + Loss_{GAN} \tag{9}$$

where $\lambda$ is the user-defined weighting factor to balance the tradeoff between two loss components. Such a form of loss function may remind the reader of the loss function of pix2pix GAN,[31] where the data-fidelity loss is replaced with a supervised $L_1$ loss. The proposed method and pix2pix GAN both try to realize the conditional image generation in the GAN framework, but the pix2pix GAN is a supervised-learning model that requires paired data for model training.

Noticeably, the correlation between the generated material images and acquired DECT images is only enforced by the DECT data-fidelity loss, and the noise suppression is enforced through the generative-adversarial mechanism.

## 3. Evaluation

### 3.1 Data acquisition

Two simulation studies and two clinical studies were conducted in this work. Single-energy CT (SECT) images were collected from 27 brain and 34 lung cases, respectively in the simulation studies, while DECT images were collected from 43 brain and 42 lung patients in the clinical cases.

For the simulation study of the brain, a total of 1426 slices from 18 patients were used for training, and 500 random slices from 9 patients were used for testing. For the simulation study of the lung, 1766 slices from 23 patients were selected for training, and 500 random slices from the remaining 11 patients were selected for testing. The head SECT data were collected from the CQ500 dataset,[32] and the lung images were collected from the LungCT-Diagnosis dataset on The Cancer Imaging Archive (TCIA). In the simulation studies, the ground-truth material-specific images were first generated from SECT using threshold and normalization, then the DECT images were generated from the material maps and associated the spectral and materials' attenuation information.[11] The details of the data generation are introduced in the next section. To demonstrate the unsupervised property of the proposed method, 1000 slices from another 12 brain patients and 1000 slices from another 15 lung patients were used for the generation of reference images, and a number of the reference material-specific images were augmented by rotation and translation to match the number of training data.

For the clinical study of the brain, 2316 slices from 32 patients were used for training and 300 random slices from 11 patients were used for testing. For the clinical study of the lung, a total of 2621 slices from 30 patients were selected for training, and 300 slices were randomly selected from 12 patients were selected for testing. There is no ground truth of decomposed images in the clinical cases. All the DECT images were acquired on Siemens TwinBeam scanner at 120 kVp with tin (Sn) and gold (Au) filters and were reconstructed in 512×512 matrix at pixel resolution of $1.0 \times 1.0$ mm$^2$. The brain images were cropped to 256×256 for computation efficiency. Similar to the simulation studies, 1000 slices from another 10 brain patients and 1000 slices from another 11 lung patients were used for reference image generation and the images were augmented for data size matching.

## 3.2 Data generation in simulation studies

In the simulation studies, only SECT images were collected and DECT images were generated through simulated scans. Firstly, the reference bone and tissue maps were obtained from CT images using the threshold method in previous research.[11] Then the dual-energy projection data were simulated using Beer-Lambert's law

$$\vec{I}_{H/L} = \int S_{H/L}(E) e^{-F(\vec{x}_1 \mu_1(E) + \vec{x}_2 \mu_2(E))} dE \tag{10}$$

where $\vec{I}$ represents the photon counts of each detector pixel, $S(E)$ is the X-ray spectrum, $F$ indicates the forward projection operator in CT implemented using Siddon's ray-tracing algorithm,[33] $\mu(E)$ is the energy-dependent linear attenuation coefficients of the basis material, subscript $H, L$ mean the high- or low-energy level, and subscripts 1,2 represent the bone and soft tissue, respectively.

In this study, the X-ray tube voltages at 140 kVp and 80 kVp with additional 1.5-mm aluminum (Al) and 0.2-mm copper (Cu) filtration are simulated with an energy resolution of 1 keV using Boone's method.[34] The energy-dependent attenuation coefficients were generated from the Spektr toolbox.[35] The scanning was in fan-beam geometry with a source-to-detector distance (SID) of 1500 mm and the source-to-axis distance (SAD) of 1000 mm, in which 600 equiangular views were acquired over 360°. The detectors were in the size of 512 pixel × 0.776 mm for brain case and 512 pixel ×1.552 mm for lung case.

With detected photon counts at each spectrum, the sinogram data were calculated by

$$\vec{b}_{H/L} = -\ln \frac{\vec{I}^n_{H/L}}{\int S_{H/L}(E) dE} \tag{11}$$

where $\vec{b}$ is the sinogram data, the superscript $n$ denotes the Poisson noise and the unattenuated photon intensity was set at $1\times10^4$ per detector pixel. The DECT images ($\vec{\mu}_{H,L}$) were analytically reconstructed from $\vec{b}_{H/L}$ using the FBP algorithm in 256×256 matrix with each pixel at resolution of 1×1 mm$^2$. For comparison, the noiseless DECT images were also reconstructed with no Poisson noise added to the projection data.

To perform the image-domain decomposition, the linear attenuation coefficients ($\mu_{1/2,H/L}$) of the bone and soft tissue need to be measured in noiseless DECT images. In this study, these values were obtained by

$$\mu = \frac{\sum_j x_j \mu_j}{\sum_j x_j} \tag{12}$$

where $x_j$ and $\mu_j$ are the volume fraction and linear attenuation coefficient of the pixel index by $j$, and all the pixels with $x \in [0.95\ 1.05]$ were included for the measurement of $\mu$.

## 3.3 Data preparation in clinical studies

In the clinical studies, the DECT images exported from the commercial scanner were in the unit of HU. So we first converted them to the unit of mm$^{-1}$ by

$$\mu = (\frac{HU}{1000} + 1) \bar{\mu}_{water} \tag{13}$$

and $\bar{\mu}_{water}$ was the average linear attenuation coefficient of water in each spectrum calculated as

$$\bar{\mu}_{water} = \int S(E)\mu_{water}(E)dE \qquad (14)$$

where $S(E)$ denotes the X-ray spectra after Sn or Au filtration.

With the converted DECT images, $\mu_{1/2,H/L}$ were obtained by manually selecting two uniform regions of interest (ROIs) on the DECT images that contain the basis materials, then the average pixel values in the two ROIs were taken as the elements of the composition matrix ($A$), as implemented in previous works.[12]

Similar to the process in simulation studies, the reference material-specific images were obtained from the high-energy CT images using the threshold method. Such reference images were not the ground truth, but only provided the feature of low-noise images for training the discriminator.

### 3.4. Comparison to the state-of-the-art unsupervised method

To demonstrate the superiority of the proposed material decomposition method, we designed another unsupervised scheme based on the denoising diffusion probabilistic model (DDPM) for a comparison study. Diffusion models are state-of-the-art generative models and the DDPM is one implementation of diffusion models that convert a standard Gaussian distribution to the target data distribution through the Markov chain.[36]

In the forward process, an image $x_0$ in the target distribution $q(x_0)$ is transformed into Gaussian noise $x_T$ by gradually contaminating the images over $T$ steps, as shown in Figure 2(a) by the orange-colored data flow. Each Markov forward process can be expressed as:

$$q(x_t|x_{t-1}) = \mathcal{N}(x_t; \sqrt{1-\beta_t}x_{t-1}, \beta_t I) \qquad (15)$$

where $\beta_t \in (0,1)$ is the predefined variance schedule, and $\mathcal{N}(x_t; \mu, \sigma^2)$ is the Gaussian distribution with mean $\mu$ and variance $\sigma^2$. Using a reparameterization strategy of $\alpha_t := 1 - \beta_t$ and $\bar{\alpha}_t := \prod_{i=1}^{t} \alpha_i$, $x_t$ at any step $t$ can be obtained directly from the initial image $x_0$:

$$x_t = \sqrt{\bar{\alpha}_t}x_0 + \sqrt{1-\bar{\alpha}_t}\epsilon \qquad (16)$$

where $\epsilon \sim \mathcal{N}(0, I)$ and $x_T$ is an isotropic Gaussian noise when $T \to \infty$. This simplified one-step calculation of $x_t$ is employed in the training stage, as shown in Algorithm 1.1.

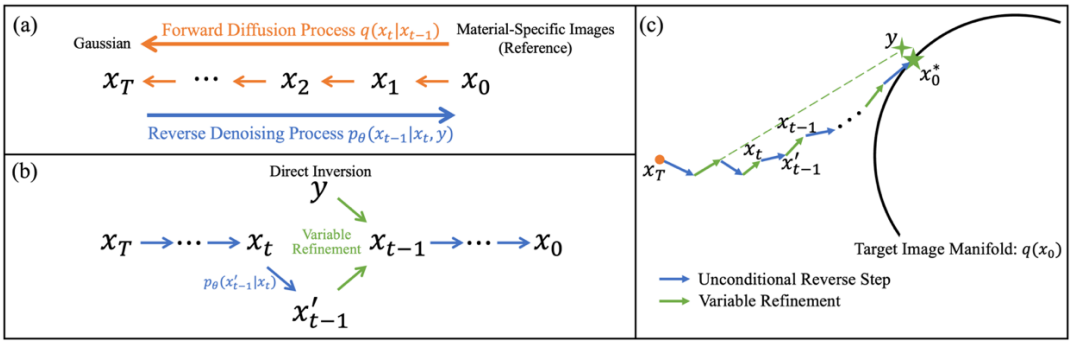

**Figure 2**. (a) Training and (b) sampling stage of the unsupervised DDPM model for material decomposition. (c) The principle of the conditional sampling process.

In the reverse process indicated by the blue-colored flow in Figure 2(a), the DDPM transforms a Gaussian noise sample $x_T$ to target image $x_0$ through

$$p_\theta(x_{t-1}|x_t) = \mathcal{N}(x_{t-1}; \frac{1}{\sqrt{\alpha_t}}(x_t - \frac{\beta_t}{\sqrt{1-\bar{\alpha}_t}}\epsilon_{\theta,t}(x_t)), \tilde{\beta}_t I) \qquad (17)$$

where $\epsilon_{\theta,t}(x_t)$ is the predicted noise from $x_t$ by a neural network with parameters of $\theta$ and $t$, and $\tilde{\beta}_t = \frac{1-\bar{\alpha}_{t-1}}{1-\bar{\alpha}_t}\beta_t$.

The original DDPM described above is an unconditional generation model without any guidance during the sampling stage (Eq. 17). However, the material decomposition is an inverse problem requiring a conditional generation model. In order to solve the inverse problem of material decomposition using DDPM in an unsupervised manner, we kept the unconditional training process unchanged and altered the sampling stage to the conditional scheme by introducing variable refinement at each step:

$$x_{t-1} = \frac{T-t+1}{T}x'_{t-1} + \frac{t-1}{T}y \qquad (18)$$

where $x'_{t-1}$ is the transition sampled from the posterior distribution (Eq. 17) and the image $y = A^{-1}[\vec{\mu}_H \quad \vec{\mu}_L]^\mathsf{T}$ is the direct inversion from DECT, as shown in Figure 2(b). The principle of the variable refinement is depicted in Figure 2(c), where $x_0^*$ is the low-noise decomposed images associated with the acquired DECT, and $y$ is inherently a noisy perturbation of $x_0^*$ which is offset from the target manifold but close to $x_0^*$. With the guidance of $y$, the variable refinement directs the generation process toward the target image $x_0^*$. The pseudo code for the conditional sampling stage is summarized in Algorithm 1.2.

**Algorithm 1**. The training and sampling procedures of the unsupervised DDPM.

| **Algorithm 1.1** Training | **Algorithm 1.2** Sampling |
|---|---|
| 1: **repeat** | 1: $x_T \sim \mathcal{N}(0, I)$ |
| 2:    $(x_0) \sim p(x)$ | 2: **for** $t = T, \cdots, 1$ **do** |
| 3:    $t \sim U([0,1])$ | 3:    $z, \epsilon \sim \mathcal{N}(0, I)$ if $t > 1$, else $z = 0$ |
| 4:    $\epsilon \sim \mathcal{N}(0, I)$ | 4:    $x'_{t-1} = \frac{1}{\sqrt{\alpha_t}}\left(x_t - \frac{1-\alpha_t}{\sqrt{1-\bar{\alpha}_t}}\epsilon_{\theta,t}(x_t)\right) + \sqrt{\tilde{\beta}_t}z$ |
| 5:    $x_t = \sqrt{\bar{\alpha}_t}x_0 + \sqrt{1-\bar{\alpha}_t}\epsilon$ | 5:    $x_{t-1} = \frac{t-1}{T}y + \frac{T-t+1}{T}x'_{t-1}$ |
| 6:    Take an optimization step on $\nabla_\theta \|\epsilon_{\theta,t}(x_t) - \epsilon\|^2$ | 6: **end for** |
| 7: **until** converged | 7: **return** $x_0$ |

Different from the proposed GAN-based framework, no acquired DECT images is used in the training stage of DDPM. Thus, the augmented images from an additional 1000 slices are no longer needed in the model training of DDPM.

### 3.5 Metrics for image quality assessment

The proposed method was compared with three other image-domain decomposition algorithms, including the direct matrix inversion, iterative image-domain decomposition with TV regularization, and the unsupervised DDPM method. The performance of the material decomposition methods was evaluated using the standard deviation (SD) and mean squared error (MSE) gauged in designated ROIs in the decomposed images, and the structural similarity index measure (SSIM) between the decomposed images and ground-truth images, and the electron density as well:

$$SD = \sqrt{\frac{\sum_{i,j}^{n_x,n_y}(x(i,j) - \bar{x})^2}{n_x n_y - 1}} \quad (19)$$

where $x(i,j)$ is the value of pixel $(i,j)$, $\bar{x}$ is the mean pixel value inside the ROI, $n_x n_y$ is the total number of pixels included in the calculation.

$$MSE = \frac{1}{n_x n_y} \sum_{i,j}^{n_x,n_y} (x(i,j) - x_{ref}(i,j))^2 \quad (20)$$

where $x_{ref}(i,j)$ is the value of pixel $(i,j)$ on the reference material-specific image. The SSIM quantified the structural preserving of the decomposed images, which was calculated as

$$SSIM = \frac{(2\bar{x}\bar{x}_{ref} + C_1)(2\sigma_{x,x_{ref}} + C_2)}{(\bar{x}^2 + \bar{x}_{ref}^2 + C_1)(\sigma_x^2 + \sigma_{x_{ref}}^2 + C_2)} \quad (21)$$

where $C_1 = (0.01 \cdot L)^2$ and $C_2 = (0.03 \cdot L)^2$ are small constants to avoid divergence caused by the division operation, and $L$ is the dynamic range of the input images. $\sigma_x^2$ and $\sigma_{x_{ref}}^2$ are the variance of decomposed and reference images, and $\sigma_{x,x_{ref}}$ is the covariance of decomposed and reference images. Electron density is calculated by

$$\rho_e = \rho_e^b \cdot \bar{x}_b + \rho_e^t \cdot \bar{x}_t \quad (22)$$

where $\bar{x}_b$ and $\bar{x}_t$ are mean values inside the ROI on decomposed bone and tissue images, and $\rho_e^b$ and $\rho_e^t$ are the electron densities of the basis materials. In this study, $\rho_e^b$ and $\rho_e^t$ were set to 4.862×10²³ and 3.483×10²³ e/cm³, referring to the manual of Gammex electron density phantom.

### 3.5 Implementation details

The proposed GAN-based material decomposition method and the DDPM model were implemented using PyTorch 1.13.1 on a 40GB Nvidia A100 GPU.

For the GAN-based method, a U-net with residual blocks was used for the generator and a PatchGAN was used for the discriminator.[37] The weighting factor of data-fidelity loss was set to 1×10⁴ in the simulation studies and 1×10² in the clinical studies. The Adam optimizer was used with the learning rate of 1×10⁻⁵ and betas of 0.5 and 0.999, respectively. The batch size was set to 2 for all the studies. The training was stopped after 200 epochs, which took about 2 h for the simulation studies, 3.5 h for the clinical study of the brain, and 10 h for the clinical study of the lung.

For the unsupervised DDPM model, a U-net structure with attention modules and residual blocks was used for noise prediction at each time step.[36] For all the studies, the total number of time steps was set to 1000 and the noise variance was linearly scheduled from 1×10⁻⁴ to 2×10⁻². The Adam optimizer was used with the learning rate of 1×10⁻⁴ and betas of 0.9 and 0.999, respectively. The batch size was set to 1 for the clinical study of the lung and 2 for other studies. The training was stopped after 1000 epochs, which took about 50 h for the simulation studies, 90 h for the clinical study of the brain, and 460 h for the clinical study of the lung.

The TV-regularized iterative decomposition was performed using the fast iterative shrinkage-thresholding algorithm (FISTA),[38] and the weighting factors of both TV terms were set to $5\times10^{-7}$ for the simulation study of the brain, $7\times10^{-7}$ for the simulation study of the lung, $9\times10^{-8}$ for the clinical study of the brain, and $1.5\times10^{-7}$ for the clinical study of the lung.

## 4. Results

### 4.1 Simulation studies

Figure 3 summarizes three slices of DECT and corresponding decomposed results using different methods in the simulation head study. The first two columns of the decomposed images are direct inversions from DECT without and with Poisson noise in the simulated scan, respectively. Consistent with previous research, matrix inversion from noisy DECT significantly increased the noise levels and severely degraded the SNRs. The TV-regularized iterative decomposition efficiently suppressed the noise in decomposed images but led to coarser noise and blurring of fine structures, as indicated by the red arrows in Figure 3. The unsupervised DDPM generated material-specific images with lower noise levels than the results of iterative decomposition. However, as indicated by the red arrows, residual noise remained in the decomposed images, because the guided images were too noisy to provide detailed structures for the image sampling. Compared to the iterative and DDPM methods, the proposed method significantly suppressed the noise in decomposed images without blurring the image. The last column lists the ground truth of the material-specific images for reference, where the red circle or box indicates a uniform ROI for SD measurement.

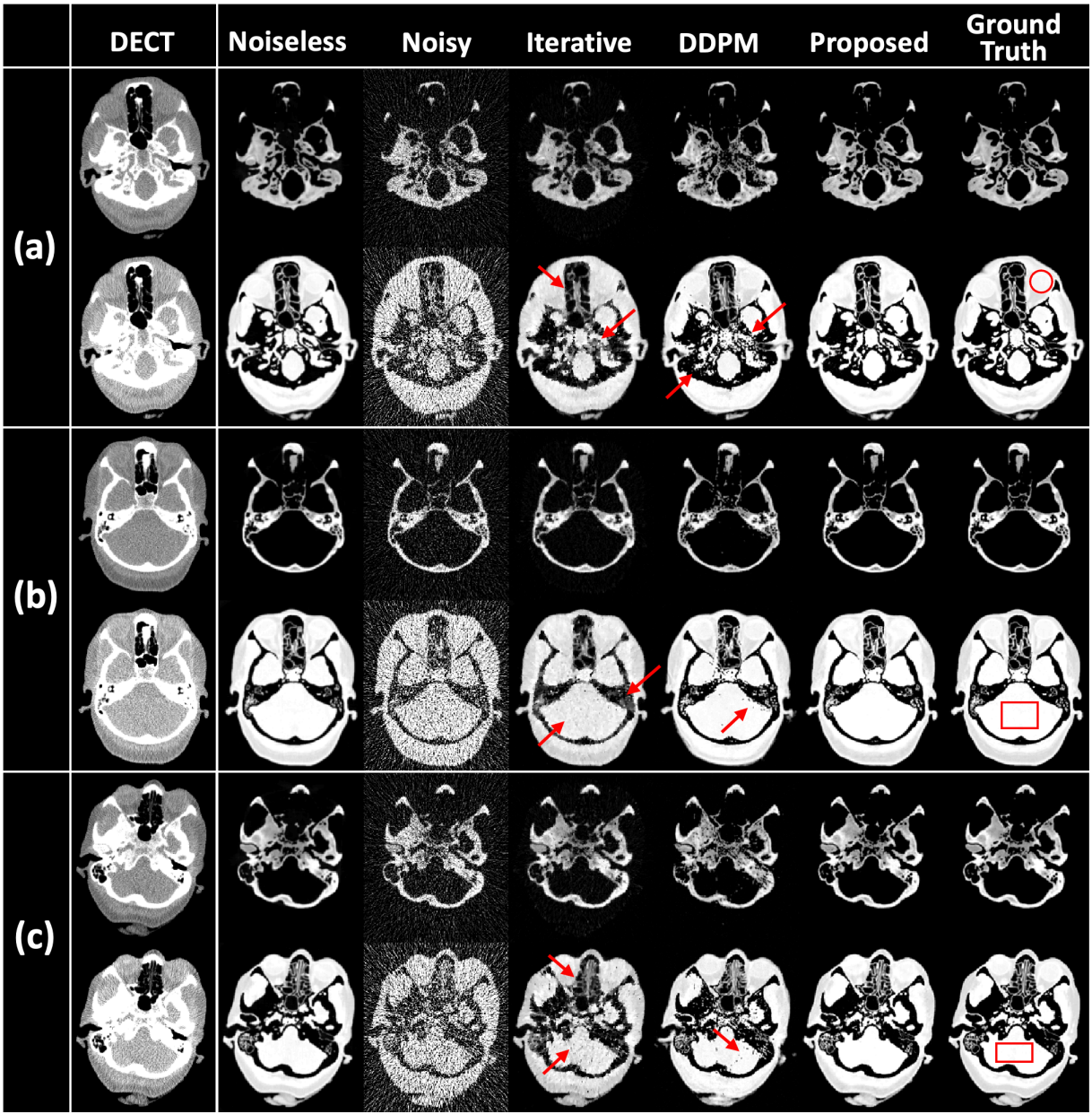

**Figure 3**. Results of the simulation brain study. The red circle or box indicates the uniform ROI for SD calculation. Display windows are [-500 500] HU for DECT and [0 1] for material-specific images.

The calculated electron density maps from the decomposed bone and tissue maps are summarized in Figure 4. The noise level and image texture of the electron density maps are consistent with the decomposed images in Figure 3. Compared to the other three methods, the proposed method produced results at the lowest noise level and best structural preserving performance.

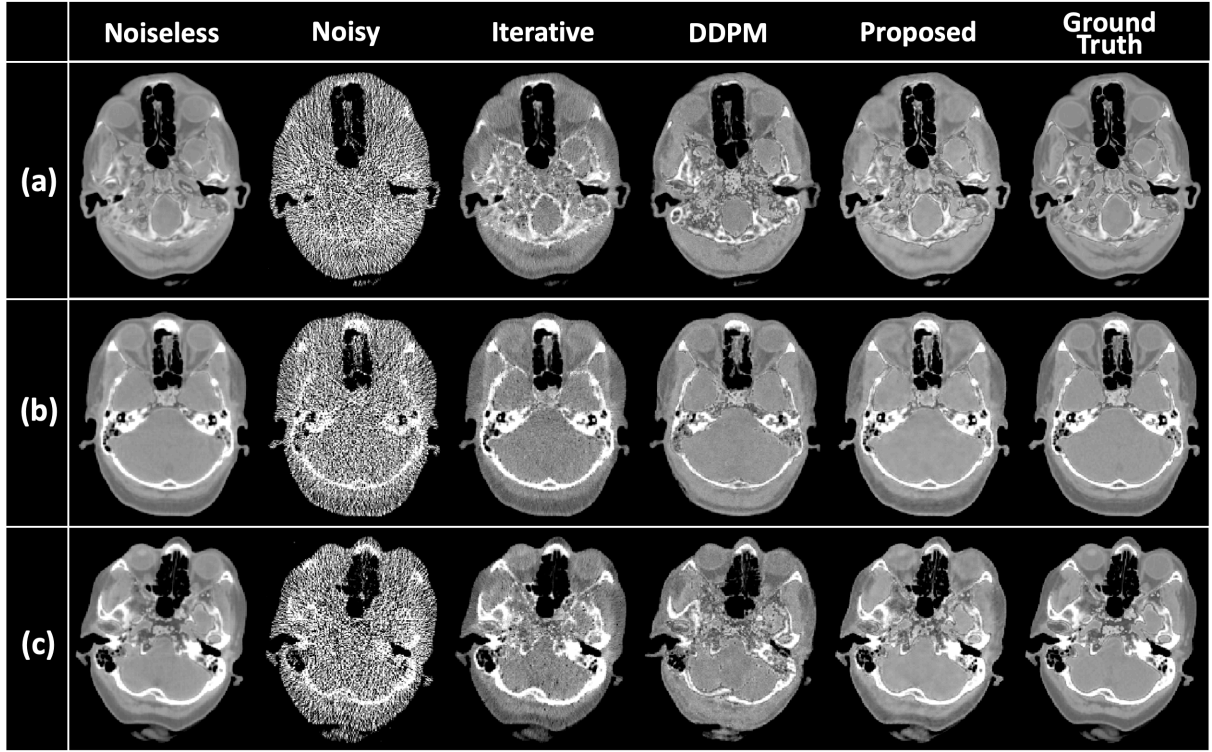

**Figure 4**. Calculated electron density maps of the slices in Figure 3. The display window is [2 4.5]×$10^6$ e/cm$^3$.

Similar results were seen in the lung study, as shown in Figures 5 and 6. The structural distortion appeared in the tissue maps generated from the DDPM method, as indicated by the red arrows in Figure 5. This is because the original structural information is lost in the direct inversion images, and then correct guidance for such structures is missing in the sampling stage, which is consistent with previous research on the DDPM-based medical image translation.[39] As it in the head study, the red circle or box indicates the ROI for noise measurement. In Figure 6, similar distortions were also observed in the electron density maps calculated from DDPM results.

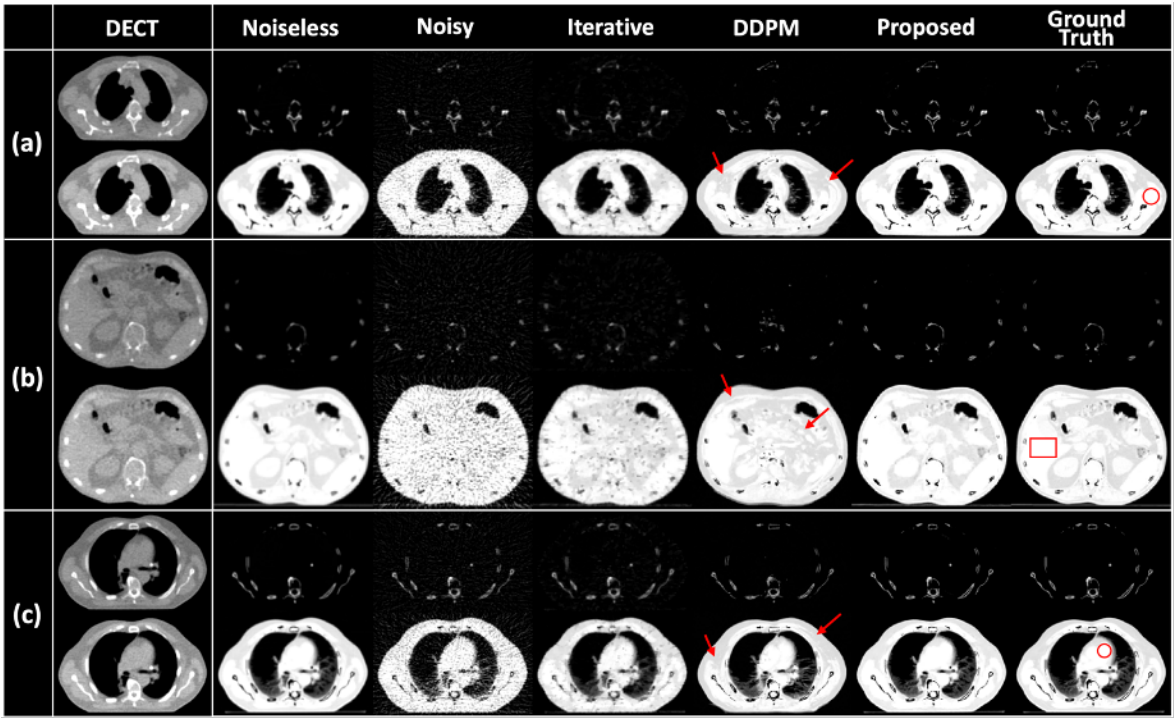

**Figure 5**. Results of the simulation lung study. The red box indicates the uniform ROI for noise SD calculation. Display windows are [-500 500] HU for DECT and [0 1] for material-specific images.

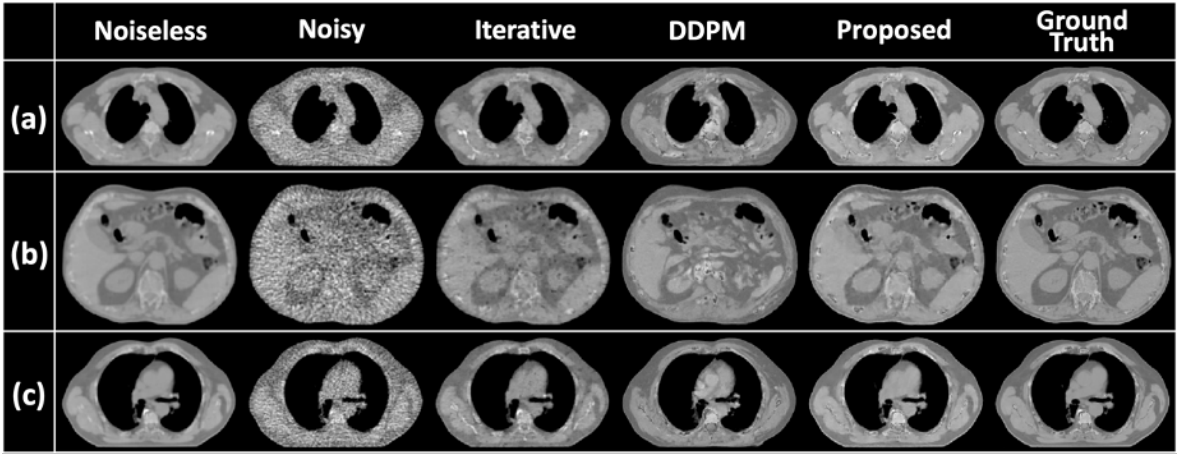

**Figure 6**. Calculated electron density maps of the slices in Figure 5. Display window is $[2\ 4.5]\times 10^6$ e/cm$^3$.

 The mean and SD gauged in the indicated ROIs are listed in Table I. Compared to direct matrix inversion, the proposed method reduced the SD by more than 97% in the brain digital phantom study, and more than 91% in the lung digital phantom study. The proposed method also outperformed the iterative algorithm and DDPM in noise suppression in both studies. The mean values of the iterative decomposition results are lower than the direct inversion because the $l_1$-norm instead of $l_0$-norm is used in the optimization, which not only enforces the signal sparsity but also reduces the signal magnitude. Lower mean values also appear in some of the DDPM results due to the residual noise or distortions inside ROIs. The results of the proposed method have almost the same mean values as the results of matrix inversion, verifying the feasibility of accurate material decomposition enabled by the introduced data-fidelity loss.

Table I. Mean values and standard deviations measured inside ROIs in Figures 3 and 5.

| (MEAN±SD) in ROI | | Direct Inversion (Noiseless) | Direct Inversion (Noisy) | Iterative Decomposition | Unsupervised DDPM | Proposed |
|---|---|---|---|---|---|---|
| Brain Study | Slice (a) | 0.98±0.01 | 0.97±0.43 | 0.91±0.03 | 0.95±0.02 | 0.99±0.01 |
| | Slice (b) | 1.01±0.01 | 1.01±0.52 | 0.92±0.04 | 0.99±0.02 | 1.01±0.01 |
| | Slice (c) | 0.99±0.01 | 0.98±0.67 | 0.87±0.09 | 0.96±0.07 | 1.00±0.01 |
| Lung Study | Slice (a) | 1.03±0.00 | 1.03±0.15 | 0.98±0.04 | 1.01±0.02 | 1.04±0.00 |
| | Slice (b) | 1.04±0.01 | 1.04±0.16 | 1.00±0.04 | 1.03±0.01 | 1.04±0.01 |
| | Slice (c) | 1.04±0.01 | 1.04±0.12 | 1.00±0.03 | 0.97±0.03 | 1.05±0.01 |

The SSIM between the decomposed images and the ground truth are summarized in Table II. Compared to the direct inversion, all the other three methods improve the SSIM while the proposed method produces the highest SSIM – a jump from roughly 0.2 to above 0.9.

Table II. Mean values of SSIM between the decomposed images and the ground truth.

| MEAN of SSIM | | Direct Inversion (Noiseless) | Direct Inversion (Noisy) | Iterative Decomposition | Unsupervised DDPM | Proposed |
|---|---|---|---|---|---|---|
| Brain Study | Bone Map | 0.835 | 0.111 | 0.601 | 0.877 | 0.990 |
| | Tissue Map | 0.787 | 0.084 | 0.776 | 0.821 | 0.962 |
| Lung Study | Bone Map | 0.899 | 0.184 | 0.698 | 0.877 | 0.987 |
| | Tissue Map | 0.937 | 0.152 | 0.811 | 0.705 | 0.951 |

With the ground-truth material-specific images in the simulation studies, the accuracy of material decomposition can be further evaluated by the error analysis of decomposed images, as listed in Table III. In the brain digital phantom study, the MSE between decomposed images and the ground truth was significantly decreased by the proposed method from 0.019 to $0.289 \times 10^{-3}$ for the bone map, and from 0.114 to $1.363 \times 10^{-3}$ for the tissue map. In the lung digital phantom study, the MSE decreased from 0.004 to $0.159 \times 10^{-3}$ and from 0.026 to $1.211 \times 10^{-3}$ for the bone and tissue maps, respectively. The proposed method outperforms the iterative algorithm and DDPM in the accuracy of material decomposition, and the accuracy of the proposed method is comparable to the noiseless case, where the errors are from the reconstruction and beam-hardening effect.

Table III. MSE of the decomposed bone and tissue images.

| MSE | | Direct Inversion (Noiseless) | Direct Inversion (Noisy) | Iterative Decomposition | Unsupervised DDPM | Proposed |
|---|---|---|---|---|---|---|
| Brain Study | Bone Map | $0.271 \times 10^{-3}$ | 0.019 | $1.310 \times 10^{-3}$ | $2.117 \times 10^{-3}$ | $0.289 \times 10^{-3}$ |
| | Tissue Map | $0.907 \times 10^{-3}$ | 0.114 | $6.569 \times 10^{-3}$ | $5.285 \times 10^{-3}$ | $1.363 \times 10^{-3}$ |
| Lung Study | Bone Map | $0.281 \times 10^{-3}$ | 0.004 | $0.934 \times 10^{-3}$ | $0.745 \times 10^{-3}$ | $0.159 \times 10^{-3}$ |
| | Tissue Map | $1.402 \times 10^{-3}$ | 0.026 | $5.248 \times 10^{-3}$ | $4.483 \times 10^{-3}$ | $1.211 \times 10^{-3}$ |

The electron density values calculated in the uniform ROIs are listed in Table IV. Consistent with the mean values in Table I, the iterative method decreases the signal intensity and leads to lower electron density values. Errors of some DDPM results come from the structural distortions in the decomposed images, and the errors of noiseless results come from the reconstruction and beam-hardening effect. The proposed method produces more accurate electron density results than iterative and DDPM methods with better noise suppression performance.

Table IV. Electron density calculation inside ROIs in Figures 3 and 5.

| Electron Density ($\times 10^{23}$ e/cm$^3$) | | Reference | Direct Inversion (Noiseless) | Direct Inversion (Noisy) | Iterative Decomposition | Unsupervised DDPM | Proposed |
|---|---|---|---|---|---|---|---|
| Brain Study | Slice (a) | 3.40 | 3.40 | 3.39 | 3.30 | 3.37 | 3.42 |
| | Slice (b) | 3.51 | 3.48 | 3.47 | 3.34 | 3.45 | 3.50 |
| | Slice (c) | 3.50 | 3.46 | 3.45 | 3.29 | 3.41 | 3.51 |
| Lung Study | Slice (a) | 3.54 | 3.60 | 3.58 | 3.52 | 3.38 | 3.63 |
| | Slice (b) | 3.56 | 3.60 | 3.60 | 3.54 | 3.58 | 3.61 |
| | Slice (c) | 3.52 | 3.54 | 3.55 | 3.47 | 3.53 | 3.56 |

## 4.2 Clinical studies

Figure 7 summarizes the results of the clinical brain study. The results of direct inversion were less noisy than the ones in the simulation studies because the clinical DECT images had lower noise levels than the simulated DECT. Similar to the simulation studies, the iterative decomposition not only suppressed the noise but also decreased the pixel values and introduced coarser noise in the decomposed images. Compared to the simulation studies, there was less remaining noise in the DDPM results because the direct inversion results for variable refinement were less noisy. The proposed method showed the best performance in noise suppression and structural preserving.

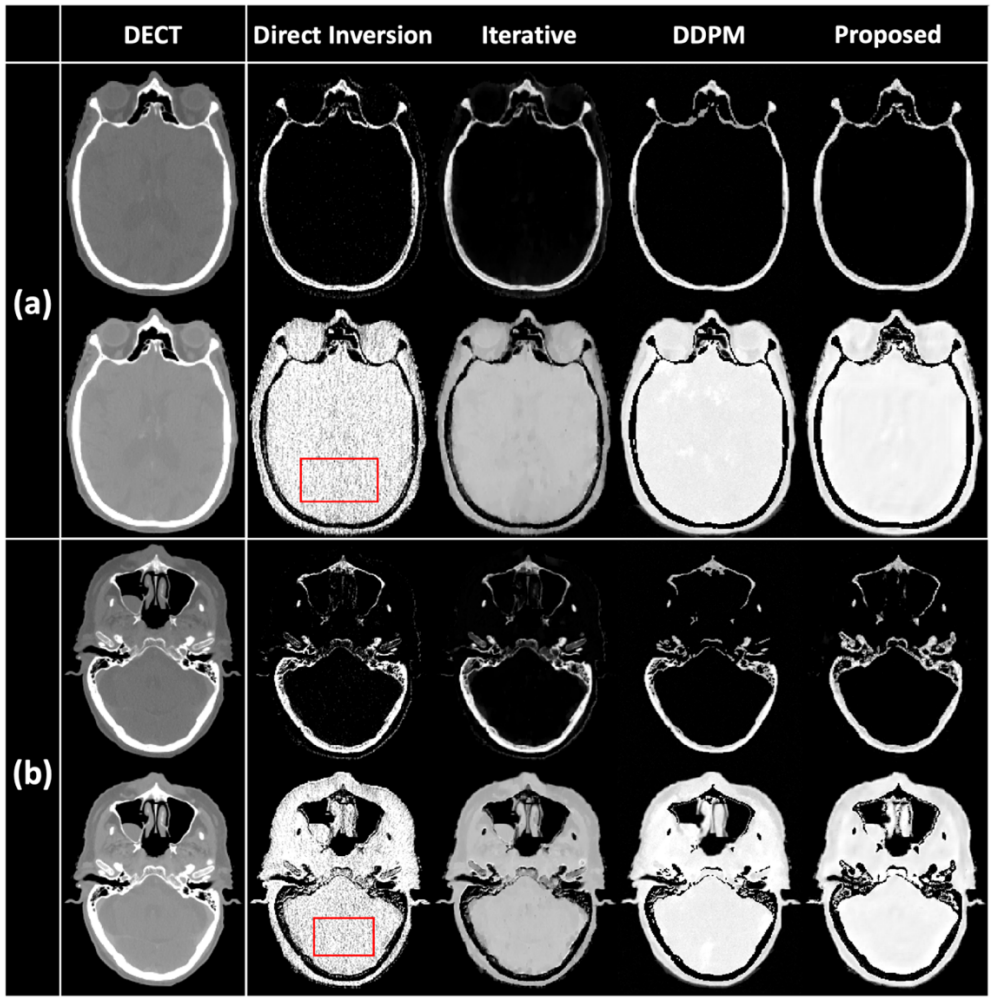

**Figure 7**. Selected slices of the clinical brain patient study. The red box indicates the uniform ROI for noise SD calculation. Display windows are [-500 500] HU for DECT and [0 1] for material-specific images.

The mean and noise gauged in the ROIs indicated in Figure 7 are listed in Table V. The proposed method significantly decreased the SD by more than 95% from the direct inversion results, showing the superiority of the proposed algorithm.

**Table V**. Mean values and standard deviations measured inside ROIs in Figure 7.

| (MEAN±SD) in ROI | Direct Inversion | Iterative Decomposition | Unsupervised DDPM | Proposed |
|---|---|---|---|---|
| Slice (a) | 1.00±0.28 | 0.79±0.05 | 0.97±0.02 | 0.97±0.01 |
| Slice (b) | 0.96±0.23 | 0.69±0.03 | 0.95±0.03 | 0.95±0.01 |

The results of the clinical lung study are shown in Figure 8, where the noise level is higher than in the brain study due to the larger object size. In this study, besides the coarser noise introduced by TV regularization, horizontal beam-hardening artifacts were also observed in the decomposed images. Similar to the simulation studies, residual noise was observed in the DDPM results due to the severely noised images from direct inversion. The proposed method generated low-noise material-specific images with sharp edges as in all other studies.

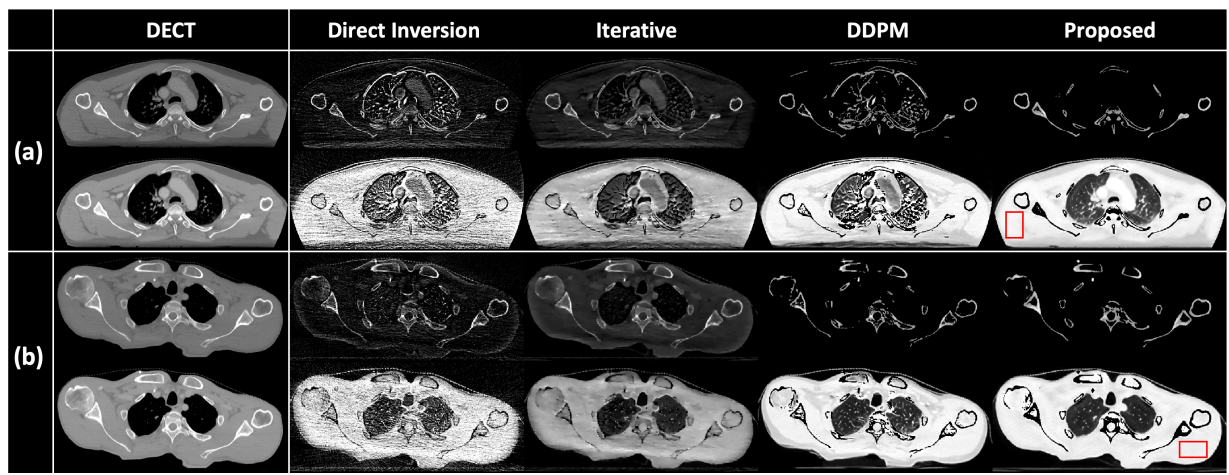

**Figure 8**. Selected slices of the clinical lung patient study. The red box indicates the uniform ROI for noise SD calculation. Display windows are [-500 500] HU for DECT and [0 1] for material-specific images.

The mean and noise gauged in the ROIs indicated in Figure 8 are summarized in Table V. All the three methods suppressed the noise while the proposed method produced the most significant noise reduction, reducing the noise by more than 93% compared to the direct matrix inversion.

**Table VI**. Mean values and standard deviations measured inside ROIs in Figure 8.

| (MEAN±SD) in ROI | Direct Inversion | Iterative Decomposition | Unsupervised DDPM | Proposed |
|---|---|---|---|---|
| Slice (a) | 0.93±0.16 | 0.73±0.02 | 0.93±0.02 | 0.94±0.01 |
| Slice (b) | 0.97±0.19 | 0.75±0.02 | 0.96±0.02 | 0.95±0.01 |

## 5. Discussion

This work formulated the DECT material decomposition under the unsupervised GAN framework. The generator is trained to carry out the image-domain material decomposition, with the data-fidelity introduced to guarantee the accuracy of material decomposition. The discriminator is trained to differentiate the generated material-specific images from the referenced images, which works in a manner similar to the image-prior regularization in iterative methods.

The significance of this work is two-fold. First, the introduction of data-fidelity loss enables GAN to solve inverse problems with an implicit image prior learned by the discriminator. By replacing the matrix $A$ with other forward operators, the proposed scheme is ready to solve other inverse problems such as inverse planning in radiotherapy, fast MRI image reconstruction, sparse-view or limited-angle CT reconstruction, metal artifact reduction, and so on. Second, the proposed unsupervised-learning method does not require paired data for model training, which has great potential of addressing the challenges due to lack of ground-truth data in clinical practice. Furthermore, the variable refinement approach in the compared DDPM method is also a novel technique, which makes it the first time to solve the inverse problem of material decomposition using the unsupervised DDPM.

The primary issue of image-domain decomposition is the inevitable beam-hardening effect. In this work, only the slices without metal implants were included to reduce the beam-hardening effect. However, moderate beam-hardening effects are still introduced by the logarithm operator in Eq. (11). In other words, one cannot obtain the ground-truth material maps even if the DECT scans are noiseless. This results in the errors of decomposed images in the noiseless simulation cases. The proposed method can be modified to correct the beam-hardening effect during material decomposition, by replacing the image-domain data-fidelity loss with the nonlinear data-fidelity loss function in the one-step projection-domain iterative decomposition algorithms. In short, the presented work is the image-domain version of the proposed framework and can be readily extended to be implemented in the projection domain. This will be the focus of our research following this work.

The other direction of future study is to further improve the unsupervised DDPM method designed for comparison studies in this work. Compared to the iterative method, the DDPM has been demonstrated to be more effective in noise suppression and structure preservation in the material decomposition. However, the DDPM results are corrupted by residual noise and distortions because the image guidance for variable refinement is too noisy to provide fine structural information. One potential solution to this issue is to incorporate image conditions with more structural information, e.g., the DECT images, into the sampling process. Another potential solution is to combine the DDPM and iterative decomposition methods. For instance, the outcome of DDPM can be fed into the iterative algorithms as an initialization, and the residual noise and distortions may be corrected during the optimization process.

There is an alternative strategy to combine the iterative decomposition and the deep learning-based image prior under the framework of unsupervised learning. For example, one can train a scoring network to recognize various noise levels in the material-specific images. Then the trained model can be embedded into the objective function of the iterative decomposition algorithm as a regularization term. The optimization problem with deep image regularization can be solved using the gradient-based algorithms. A similar strategy has been implemented for metal artifact reduction in CT.[40] Noticeably, there are two major differences between this deep prior-regularized iterative decomposition scheme and the proposed method. The first is that the basis materials have been prefixed during the neural network training in the proposed GAN-based method, while the basis materials can be user-defined during the optimization in the alternative strategy. The second is that the generation of decomposed images in the proposed method is carried out in a single step without any iterations, whereas the alternative strategy needs to solve the complicated optimization problem every time. There is a trade-off between the freedom of parameter selection and the cost of computation for material decomposition, and one may select the strategy customed for the specific task to be accomplished.

As presented in the tissue maps in Figures 7 and 8, the proposed method showed limited ability in low-contrast tissue differentiation, which is because the energy separation between the Au- and Sn-filtered spectra is limited in the TwinBeam scanner. This may result in potential limitations for diagnostic imaging tasks, but the decomposed images generated by the proposed method can fully meet the requirement of dose calculation in radiation therapy.

## 6. Conclusions

In this work, we proposed an image-domain material decomposition method for DECT in an unsupervised-learning framework, which does not require paired DECT and ground-truth material-specific images for model training. The proposed method produces accurate decomposed images with efficient noise suppression, allowing further quantitative applications of DECT in clinical practice.

**Acknowledgments**

This research is supported in part by the National Institutes of Health under Award Number R01CA215718, R01CA272991, and P30CA008748.